\documentstyle[prd,aps]{revtex}
\begin{document}
\draft
\title{CHARGE~RENORMALIZATION~IN~A~NEW~KIND~OF~NON-LOCAL
QUANTUM~ELECTRODYNAMICS}
\author{S.~S.~Sannikov}
\address{Physico-Technical Institute\\
1 Academichna St., 310108 Kharkov, {\bf UKRAINE}}
\author{A.~A.~Stanislavsky}
\address{Institute of Radio Astronomy of the
Ukrainian National Academy of Sciences\\
4 Chervonopraporna St., 310002 Kharkov, {\bf UKRAINE}\\
E-mail: stepkin.@ira.kharkov.ua}
\date{\today}
\maketitle
\begin{abstract}
      The goal of this message is to calculate radiative corrections
to the Sommerfeld fine structure constant in the
framework of a new QED in which particles are described by bilocal
fields. The bare constant is 1/136 where 136 is a
dimension of the dynamical group of the bihamiltonian system
underlying the suggested elementary particle theory. Our
calculations in the second order of perturbation theory give the
renormalized Sommerfeld constant 1/137.0345. We believe the
difference (137.0359 - 137.0345) between corresponding
experimental and theoretical values may be understood as
corrections of the fourth order.
\end{abstract}

\section{Introduction}
      The aim of this paper is to show how to calculate the main
radiative corrections in quantum electrodynamics improved on the
bases of the general elementary particle theory suggested in \cite{1}.
The keystone of the theory is the assumption that the true
mechanism of production of elementary particles is not interactions
between them (or between their hypothetical constituents),
but is a certain quantum-dynamical system determining the special
physics at supersmall distances where the space-time is discontinuum,
i.\ e.\ it is the quite non-connected manifold there. The transition
in such a system lead to the creation of fundamental particle fields
which are bilocal wave functions $\psi(X,Y)$ in our theory (the
Heisenberg-Schr$\rm\ddot o$dinger-Dirac theory postulates the
existance of local fields $\psi(X)$, but in that theory there are
ultraviolet devergences). The initial principles of our approach
to the elementary particle problem have been stated in the Russian
periodicals \cite{1,2,3,4,5}.

      The dynamical system mentioned above has been described in
\cite{2} and named as a relativistic bi-Hamiltonian system. Owing
to the discontinuity of space in small, the quantum theory of the
system is non-unitary; the non-standard (non-Fock) representations
of the Heisenberg algebra $h^{(*)}_{16}$ described in \cite{4}
(extraction of square root of Dirac-Grassmann spinors leads
to such algebras \cite{3}) and non-unitary (infinite-dimensional)
representations of the rotation group $SO(3)$ and Lorentz group
$SO(3,1)$, induced by them and charaterized by the arbitrary complex
spin found in \cite{5} earlier, form the mathematical foundation of
this theory. Thus these representations stand for a new physical
reality. The elementary particle theory based on them is more
like the atomic spectrum theory rather than any composite model.

      In the framework of this theory (quantum electrodynamics with
bilocal fields) we consider here only one question --- the charge
renormalization which was not solved up till now.

\section{Bilocal fields and their interactions}
      The field bilocality $\psi(X,Y)$ is the direct consequence
of the semispinor structure of the particle fields $\psi^\Sigma
\sim\langle\dot f,O^\Sigma f\rangle$($O^\Sigma$ is elements
of the Heisenberg algebra $h^{(*)}_{16}$) discovered by means of
extracting the square root of Grassmann spinors, see \cite{3} (this
structure is quite analogous to the spinor structure of current
$j_\mu\sim\bar\psi\gamma_\mu\psi$ where $\gamma_\mu$ are
elements of the Clifford algebra discovered by means of extracting
the Dirac square root of vectors).

      The bilocal field $\psi^\Sigma(X,Y)$ defined by the transition
amplitude $\langle\dot f(X-Y),O^\Sigma f(X+Y)\rangle$ where
$O^\Sigma f(x)$ and $\dot f(\dot x)$ are the initial (excited) and
final (ground) states of the relativistic bi-Hamiltonian system
respectively (the explicit form of these states found in \cite{1})
is written down as
\begin{equation}
\psi(X,Y)=\frac{1}{(2\pi)^{3/2}}\int e^{ipX+iqY}\theta(p_0+q_0)\,
\theta(p_0-q_0)\,\delta(p^2+q^2)\,\delta(pq)\,\delta(p^2-m^2)\,
\psi(p,q)\,d^4p\,\frac{d^4q}{2\pi}\ .
\label{eq1}
\end{equation}
Here $X_\mu$ are coordinates in Minkowsky space ($p_\mu$ is a
4-momentum of a particle) and $Y_\mu$ are internal coordinates
(which are not fixed in the experiment and therefore we call them
hidden) describing the space-time structure of particles ($q_\mu$
is a 4-momentum of tachyon; it is interesting to note that
analogous objects have already been introduced by Yukawa \cite{6}).

      It follows from (\ref{eq1}) that if $\vert X\vert\gg\vert Y
\vert$ then the bilocal field $\psi(X,Y)$ transforms into the
usual local field $\psi(X)=\psi(X,0)$ (hence, in the suggested
scheme the local fields appear as asymptotic fields; it is a
principal point of a new correspondence principle). It also
follows from (\ref{eq1}) that $\psi(X,Y)$ may be represented in the
form of $\psi(X,Y)=F(Y,-i\frac{\partial}{\partial X})\psi(X)$
where $\psi(X)$ is a local field and $F$ is the so-called
smearing operator which has the form in the case of massive
particles ($p_\mu$ is a 4-momentum of such a particle)
\begin{equation}
F(Y,p)=\frac{1}{2\pi}\int e^{iqY}\delta(p^2+q^2)\,
\delta(pq)\,d^4q\ .
\label{eq2}
\end{equation}

      Another form of the smearing operator takes place in the
case of massless particles (it follows from the explicit form
of the leptonic transition amlplitude; it is necessary to note
that operator (\ref{eq2}) does not transform into (\ref{eq3}) when
$p^2=0$; in the case we have the stochastic integral
\begin{displaymath}
\frac{1}{2}\int_{-1}^1 e^{i\alpha pY}\,d\alpha\,),
\end{displaymath}
namely:
\begin{equation}
F_0(Y,k)=e^{iYk}\ .
\label{eq3}
\end{equation}
It is a translation ($k_\mu$ is a 4-momentum of such a particle).
Interactions between bilocal fields are described by differential
equations in Minkowsky space. We are interested in the Dirac field
$\psi(X,Y)$ interacting with the electromagnetic field $A_\mu(X,Y')$
(a general mechanism driving interactions is described in \cite{1}).
In this case the equations are written in the form of
\begin{equation}
\left(\gamma_\mu\frac{\partial}{\partial X_\mu}+m\right)\psi(X)=
-iJ(X)
\label{eq4}
\end{equation}
where $J(X)=e\gamma_\mu\int\psi(X,Y)\,A_\mu(X,Y')\,d\mu(Y,Y')$
is the interaction ``current''. It transforms into the usual
local connection between local fields $e\gamma_\mu\psi(X)A_\mu(X)$
if $\vert X\vert\gg\vert Y\vert$ (it is the new correspondence
principle from which follows the explicit form of measure:
\begin{displaymath}
d\mu(Y,Y')=\frac{\kappa^8}{(2\pi)^4}\,e^{iYY'\kappa^2}\,d^4Y\,
d^4Y'\,.
\end{displaymath}
Here $\kappa$ is a new fundamental constant equal to $\kappa=
5\cdot 10^{13}cm^{-1}$, see \cite{1}; it will be convenient for
our further calculations to put $c=\hbar=\kappa=1$).

      Proceeding from (\ref{eq4}) we may construct the
$S$ - matrix: $S=T\exp(i\int\pounds_i(X)\,d^4X$) where $\pounds_i(X)
=\frac{1}{2}[\bar\phi(X)J(X)+\bar J(X)\psi(X)]$ is an interaction
Lagrangian. In the perturbation theory the interaction picture
may be described by the well-known Feynman diagrams in vertices of
which the electron-photon formfactor arises
\begin{equation}
\rho(p,k)=\int F(Y,p)\,F_0(Y',k)\,d\mu(Y,Y')=
\frac{1}{2\pi}\int e^{iqk}\delta(p^2+q^2)\,
\delta(pq)\,d^4q\ .
\label{eq5}
\end{equation}

\section{The main formula}\label{kd}
      First of all we state the result of our calculations of
radiative corrections to the Sommerfeld fine structure constant
$\alpha=e^2/4\pi$. In the suggested theory the renormalized constant
$\tilde\alpha$ connects with the ``bare'' constant $\alpha$ by the
formula
\begin{equation}
\tilde\alpha=\left(\frac{Z_1}{Z_2}\right)^{2}\frac{Z_4}{Z_3}\,\alpha
\label{eq6}
\end{equation}
where $Z_1,Z_2,Z_3,Z_4$ are the renormalization constants of the
fermion Green function, vertex function, Lagrangian of classical
electromagnetic field and three-tail, respectively. Here all these
quantities are calculated in the second order of perturbation theory.

      In the suggested theory the ``bare'' constant $\alpha$ is
equal to 1/136 (Eddington formula) where 136 is the dimension of the
dynamical group (the group of automorphisms $Sp^{(*)}(8,{\bf C})$
for the Heisenberg algebra $h_{16}^{(*)}$) in our relativistic
bi-Hamiltonian system, see \cite{2}.

      We see formula (\ref{eq6}) essentially differs from the local
theory formula $\tilde\alpha=Z_3^{-1}\alpha$ \cite{7} being a
consequence of the Ward identity $Z_1=Z_2$ (in this theorem
for regularized constants, see \cite{7}, the regularized fermion
self-energy operator $\Sigma(p)$ is assumed to be an analytic
function at point $p^2=0$; but it does not take place in the
suggested theory: it follows from \cite{8} that $\Sigma(p)\sim
\ln p^2$ when $p^2\to 0$) and also the Furry theorem (which does not
take place in the suggested theory too due to the presence of
hidden parameters $Y_\mu$ for bare particles and the absence of them
for bare antiparticles, see further).

\section{Calculation of $Z_1/Z_2$}
      In our theory the Ward identity
\begin{displaymath}
\frac{\partial\Sigma(p)}{\partial p_\mu}+\Lambda_\mu(p,0)=0
\end{displaymath}
($\Lambda_\mu$ is the vertex function) is replaced by a more
general identity
\begin{equation}
\frac{\partial\Sigma(p)}{\partial p_\mu}+\Lambda_\mu(p,0)=
\Sigma_\mu(p)
\label{eq7}
\end{equation}
where $\Sigma_\mu(p)$ is the following operator
\begin{displaymath}
\Sigma_\mu(p)=\frac{e^2}{i(2\pi)^4}\int\frac{\gamma_\nu(\hat p-
\hat k+m)\gamma_\nu}{[(p-k)^2-m^2]\,k^2}\,\left[\frac{\partial}
{\partial p_\mu}\rho(p,k)\right]\,d^4k=
\end{displaymath}
\begin{displaymath}
=\frac{ie^2}{(2\pi)^4}
\int^1_0 dz\int_0^\infty\frac{d\sigma}{\sigma^2}\exp[i\frac{p^2}
{2\sigma}-i\sigma(m^2z-p^2z(1-z))]\left[p_\mu(2m-\hat p(1-z))+
\frac{z}{3}(p_\mu\hat p-\gamma_\mu p^2)\right].
\end{displaymath}
To express the quantity $(Z_1/Z_2-1)$ of interest to us in terms
of $\Sigma_\mu(p)$, it is necessary to take the operator on the mass
shell $\bar p=m$ by means of the formula
\begin{displaymath}
\left(\frac{Z_1}{Z_2}\right)\gamma_\mu=\Sigma_\mu(m)=
-\gamma_\mu\frac{e^2}{4\pi^2}m^2\int^1_0 z\,(1+z)\,K_1(m^2z)\,dz
\end{displaymath}
where $K_1$ is the MacDonald function. From here we get
\begin{equation}
\frac{Z_1}{Z_2}=\cases{1-\frac{3\alpha}{2\pi}\ \ \ ,\ m\ll 1;\cr
1-\frac{\alpha}{2m^2}\ ,\ m\gg 1.\cr}
\label{eq8}
\end{equation}

\section{Calculation of $Z_4/Z_3$}
      Similarly, another Ward identity
\begin{displaymath}
\frac{\partial\Pi_{\mu\nu}(k)}{\partial k_\sigma}+
\Delta_{\mu\nu\sigma}(k,0)=0
\end{displaymath}
($\Pi_{\mu\nu}$ is the polarization tensor), see \cite{9}, is
replaced by a more general identity
\begin{equation}
\frac{\partial\Pi_{\mu\nu}(k)}{\partial k_\sigma}+
\Delta_{\mu\nu\sigma}(k,0)=\Pi_{\mu\nu\sigma}(k)
\label{eq9}
\end{equation}
where $\Pi_{\mu\nu\sigma}(k)$ is the following expression
\begin{displaymath}
\Pi_{\mu\nu\sigma}^{(1/2)}(k)=\frac{ie^2}{(2\pi)^4}2\int
\frac{2p_\mu p_\nu+2p_\mu k_\nu -\delta_{\mu\nu}(p^2+pk)}
{p^2\,(p+k)^2}\,\left[\frac{\partial}{\partial k_\sigma}
\rho\Bigl((p+k)^2\Bigr)\right]\,d^4p
\end{displaymath}
in the case of Weyl's dissociation, and
\begin{displaymath}
\Pi_{\mu\nu\sigma}^{(0)}(k)=-\frac{ie^2}{(2\pi)^4}4\int
\frac{p_\mu p_\nu+p_\mu k_\nu}
{p^2\,(p+k)^2}\,\left[\frac{\partial}{\partial k_\sigma}
\rho\Bigl((p+k)^2\Bigr)\right]\,d^4p
\end{displaymath}
in the case of Klein-Gordon's dissociation.

      Speaking about the electromagnetic wave dissociation we should
explain two points. Firstly calculating $\Pi_{\mu\nu}$ we use quite
another formfactor not (\ref{eq5}), but
\begin{equation}
\rho(p^2)=\int F(Y,p)\,F(Y',p)\,d\mu(Y,Y')=
\frac{\sin p^2}{p^2}
\label{eq10}
\end{equation}
because the Lagrangian $\hat A_\mu\hat{\bar\psi}\gamma_\mu\hat\psi$
all fields of which are quantized does not give any contribution to
the charge renormalization, see \cite{8}. Another Lagrangian, namely
$A_\mu\hat{\bar\psi}\gamma_\mu\hat\psi$ ($A_\mu$ is a classical
field), gives such a contribution. If the wave function of photons
$A_\mu(X,Y)$ has the internal variables $Y_\mu$, then the classical
one (Maxwell field $A_\mu(X)$), as an essential alloy of indefinite
number of photons (light molecule), does not have such variables.
Therefore in the case only internal variables of intermediate
particles (not antiparticles) are paired. This operation leads to
the formfactor (\ref{eq10}).

      It is important to note that the bare particles as objects
being created in the transition $f\to\dot f$ have the additional
variables $Y$. The bare antiparticles arised in consequence of
interactions do not have such variables ($T$-asymmetry of 100 per
cent or complete fermion-antifermion asymmetry of the theory,
see \cite{1}). Under these circumstances the well-known Furry
theorem is invalid.

      Secondly, the polarization tensor $\Pi_{\mu\nu}$ having a
finite value $\Pi_{\mu\nu}(k)=(k_\mu k_\nu-\delta_{\mu\nu}k^2)
\Pi(k^2)+\delta_{\mu\nu}d(k^2)$ where
\begin{displaymath}
d(k,m)=-\frac{e^2}{4\pi^2}\int^1_{-1}\frac{d\alpha}{2}\int^1_0
dz\int_0^\infty\frac{\sigma\,d\sigma}{(\sigma+\alpha)^2}\,
\left[m^2-\frac{i}{\sigma+\alpha}-k^2\frac{\sigma z}{\sigma+
\alpha}\left(1-\frac{\sigma z}{\sigma+\alpha}\right)\right]\times
\end{displaymath}
\begin{displaymath}
\times\exp\left[-i\sigma m^2+ik^2\sigma z\left(1-
\frac{\sigma z}{\sigma+\alpha}\right)\right]
\end{displaymath}
for both the Dirac and Kemmer-Duffin (or Klein-Gordon)
polarizations (the expression for $\Pi$ is not given here) must be
a gauge-invariant quantity. Therefore we require $d(k)=0$ at least
in the region $k^2=0$. The last condition leads to the equation
\begin{displaymath}
\int_0^\infty\frac{\sin x}{x+m^2}\,dx+m^2\int_0^\infty
\frac{\cos x}{x+m^2}\,dx=\frac{\pi}{2}
\end{displaymath}
which has the only solution $m=0$.

      Hence a classical electromagnetic wave may dissociate on
massless particles only. Essentially, in the suggested theory there
are two and only two charged particles with zero bare mass: positron
(in our scheme it is the fundamental fermion with spin 1/2;
electron is antifermion) and $\pi$-meson (quantum of degeneration
fields with spin 0). Therefore we consider only two these cases.

      Since $\Pi_{\mu\nu\sigma}$ leads the Lagrangian to the form
of $\Pi_{\mu\nu\sigma}(k)\,A_\mu(k)\,A_\nu(k)\,A_\sigma(0)$ and in
consequence of the Lorentz-gauge $k_\mu A_\mu(k)=k_\nu A_\nu(k)=0$
we should hold only the term $\delta_{\mu\nu}k_\sigma$ in
$\Pi_{\mu\nu\sigma}(k)$. Therefore we write $\Pi_{\mu\nu\sigma}(k)
=\delta_{\mu\nu}k_\sigma I(k)$. Our calculations give
\begin{displaymath}
I^{(1/2)}(k)=\frac{e^2}{4\pi^2}\int^1_{-1}\frac{\alpha\,d\alpha}
{2}\int^1_0 dz\int_0^\infty\frac{\sigma\,d\sigma}{(\sigma+
\alpha)^3}\,\left(\frac{1}{2}-\frac{2\sigma z}{\sigma+\alpha}\right)\,
\exp\left[ik^2\sigma z\left(1-\frac{\sigma z}{\sigma+\alpha}\right)
\right],
\end{displaymath}
\begin{displaymath}
I^{(0)}(k)=-\frac{e^2}{4\pi^2}\int^1_{-1}\frac{\alpha\,d\alpha}
{2}\int^1_0 z\,dz\int_0^\infty\frac{\sigma^2\,d\sigma}{(\sigma+
\alpha)^4}\,\exp\left[ik^2\sigma z\left(1-\frac{\sigma z}
{\sigma+\alpha}\right)\right].
\end{displaymath}
On the mass shell $k^2=0$ we get
\begin{displaymath}
I^{(1/2)}(0)=-\frac{e^2}{48\pi^2}\,,\qquad
I^{(0)}(0)=-\frac{e^2}{24\pi^2}\,.
\end{displaymath}

      The quantity ($Z_4/Z_3-1$) of interest to us is determined
by the sum $I^{(1/2)}(0)+I^{(0)}(0)$ and we have
\begin{equation}
\frac{Z_4}{Z_3}=1-\frac{\alpha}{12\pi}-\frac{\alpha}{6\pi}=
1-\frac{\alpha}{4\pi}\ .
\label{eq11}
\end{equation}

\section{The principal result}
      Expressions (\ref{eq8}) and (\ref{eq11}) together give
\begin{displaymath}
\left(\frac{Z_2}{Z_1}\right)^{2}\frac{Z_3}{Z_4}=
\left(1+\frac{3\alpha}{\pi}\right)\left(1+\frac{\alpha}{4\pi}\right)=
1+\frac{13\alpha}{4\pi}\ .
\end{displaymath}

      From (\ref{eq6}) it follows now
\begin{equation}
\tilde\alpha^{-1}=\alpha^{-1}+\frac{13}{4\pi}\ .
\label{eq12}
\end{equation}

      Since in the suggested theory $\alpha^{-1}$ = 136, the
renormalized constant $\tilde\alpha^{-1}$ is $\tilde\alpha^{-1}$
= 136+1.0345 = 137.0345. The modern experimental value of this
constant is 137.0359 \cite{10}. We believe the difference 0.0014
(indeed, 0.00085 only) may be explained by the fourth order
radiative corrections.

\section{Fermion anomalous magnetic moment}
      According to the suggested theory, calculations of the vertex
operator in the third order of the perturbation theory lead to
the following formula of the fermion anomalous magnetic moment
\begin{equation}
\Delta\mu=\frac{\alpha}{\pi}\ m^2\int^1_0 z\,(1-z)\,K_1(m^2z)\,dz\ .
\label{eq13}
\end{equation}

      a) In the case $m\ll 1$ the formula (\ref{eq13}) gives
Schwinger's result $\frac{\alpha}{2\pi}$ with a correction
\begin{displaymath}
\Delta\mu\simeq\frac{\alpha}{2\pi}\left[1+\frac{m^4}{12}
\left(C-\frac{13}{12}-\ln 2+\ln m^2\right)\right]\ .
\end{displaymath}
The electron has $m=\frac{m_e c}{\kappa h}=5\cdot 10^{-4}$ and the
correction $\frac{\alpha}{2\pi}\,\frac{m^4}{12}\left(C-\frac{13}
{12}-\ln 2+\ln m^2\right)\simeq -9.8\cdot 10^{-17}$ is far beyond
the experimental possibilities of today. The $\mu$-meson has $m_\mu=
0.1$ and the correction is equal $-5.6\cdot 10^{-8}$ within the
bounds of possibility. The correction should be added to the factor
$\left(\frac{g-2}{2}\right)_{\rm theory}=\frac{\alpha}{2\pi}+0.76
\left(\frac{\alpha}{\pi}\right)^2=$0.0011655102 calculted by means
of the local theory. Its experimental value is $\left(\frac{g-2}{2}
\right)_{\rm exper}=$0.001165923 . The difference $\left(\frac{g-2}
{2}\right)_{\rm exper}-\left(\frac{g-2}{2}\right)_{\rm theory}=$
0.000000413 (together with our correction the value is equal
0.000000493) is usually accounted for by influence of the strong
interaction the correct theory of which is known to be wanting as
yet (and all the calculations are not strictly defined). However
there is a correction close to it in magnitude because of
nonlocality (of both electromagnetic and strong interactions),
i.\ e.\ owing to the finite third fundamental constant $\kappa$.

      b) For $m\gg 1$ the formula (\ref{eq13}) gives
\begin{displaymath}
\frac{g-2}{2}\simeq\frac{\alpha}{2m^2}\ .
\end{displaymath}
Let us apply it to the $\tau$-meson having $m_\tau=$1.78 and
obtain $\frac{g-2}{2}=$0.001151584 .

\end{document}